# Thermal interface conductance between aluminum and aluminum oxide: A rigorous test of atomistic level theories

Murali Gopal Muraleedharan,[1] Kiarash Gordiz,[2] Shenghong Ju,[3] Junichiro Shiomi,[3,4] Vigor Yang,[1] and Asegun Henry[5,6,7]

[1]*School of Aerospace Engineering, Georgia Institute of Technology, Atlanta, Georgia 30332, USA;*
[2]*Colorado School of Mines, 1500 Illinois Street, Golden, CO 80401, USA;*
[3]*Department of Mechanical Engineering, The University of Tokyo, Tokyo, Japan;*
[4]*Graduate School of Frontier Sciences, The University of Tokyo, Kashiwa, Japan;*
[5]*George W. Woodruff School of Mechanical Engineering, Georgia Institute of Technology, Atlanta, GA 30332, USA;*
[6]*Heat Lab, Georgia Institute of Technology, Atlanta, GA 30332, USA;*
[7]*School of Materials Science and Engineering, Georgia Institute of Technology, Atlanta, GA 30332, USA*

**Abstract**

We report the first ever accurate theoretical prediction of thermal conductance of any material interface. Thermal interfacial conductance of aluminum (Al)-sapphire (α-$Al_2O_3$) interface along crystal directions (111) Al ∥ (0001) $Al_2O_3$ for temperature ranging from 50-500 K is calculated using two fundamentally different methods: interfacial conductance modal analysis (ICMA) and atomistic green function (AGF). While AGF overpredicts interfacial conductance, both the quantitative and qualitative predictions of ICMA are exceptional when compared with the time-domain thermoreflectance (TDTR) experimental data. The mean error in ICMA results are below 5%. We believe that the accurate theoretical prediction by ICMA can be credited to a more fundamental treatment of the interfacial heat flux in contrast to that of the *phonon gas model* (PGM) and inclusion of anharmonicity to full order. ICMA also gives the eigen mode level details revealing the nanoscale picture of heat transport: more than 90% of conductance is contributed by the cross correlation (interaction) between partially extended modes of Al and $Al_2O_3$ and the remaining is attributed to interfacial modes. This is a major milestone in combustion heat transfer research enabling materials scientists to rationally design propellant architectures to serve long-distance propulsion missions.

## 1. Introduction

Understanding the nanoscale thermal behavior of materials is crucial to myriad engineering applications like space propulsion,[1-4] electronics,[5] and biomedical engineering.[6] In this century of space race, when mars and *beyond mars* propulsion missions are not distant realities, recently developed aluminum-based nano-structured energetic composites[2,7] seem promising as propellant materials. Heat conduction to unburnt reactants is the key to efficient combustion of such materials. For a typical propellant, the interface between the metallic fuel particle and its oxide layer offers high resistance to heat flow, subsequently limiting heat conduction rates. This subdues their performance in propulsion missions eventually impeding their reliability. In this regard, understanding the physical mechanisms that lead to this high interfacial resistance is of significant value. Interfacial heat transport can be characterized by thermal interfacial conductance (TIC) (denoted by $G$), which is the inverse of thermal resistance. $G$ is the constant of proportionality in the equation that relates heat flow ($Q$) at the interface of two materials to the temperature drop ($\Delta T$) at the interface ($Q = G\Delta T$). A fundamental understanding of the governing mechanisms of $G$ enables us to reengineer interfaces by designing new architectures, doping, functionalizing, etc. to improve interfacial thermal transport.

To understand and quantify $G$, one may seek experimental measurements, theory-based predictive models, or a combination of those. Time-domain thermoreflectance (TDTR) method, an optical-pump probe technique, is the most widely used experimental method.[8-11] A typical TDTR experiment measures the total conductance but neither resolves the modal contributions nor elucidates the governing mechanisms. Sometimes, merely due to the low thermal conductivity of the constituent sides of the interface, the measurement is incapable of measuring $G$ because of the low sensitivity to the interfacial resistance. In addition, experimental methods can get challenging and expensive especially from a standpoint of making clean and defect-free interfaces by epitaxial growth, and achieving high temperatures and/or pressures. Moreover, the results are sensitive to experimental conditions thus challenging reproducibility.

Theoretical approaches, in contrast to experiments, are highly reproducible, much less expensive, and can be performed over wide range of temperatures and pressures. Nevertheless, theoretical models should be thoroughly benchmarked with experimental results before being applied to any practical application. To model interfacial heat transfer, several formalisms/models exist: acoustic mismatch model (AMM),[12,13] diffuse mismatch model (DMM),[14-16] atomistic green function (AGF),[17-19] wave packet method (WP),[20-23] harmonic lattice dynamics (LD) based approach,[24,25] and frequency-domain perfectly matched (FD-PML) method[26,27]. All of them are based on the *phonon-gas model* (PGM). According to PGM, interfacial conductance is defined via Landauer formalism[28] as:

$$G = \sum_{p_A} \left[ \frac{1}{V_A} \sum_{k_{x,A}=-k_{\max}}^{k_{\max}} \sum_{k_{y,A}=-k_{\max}}^{k_{\max}} \sum_{k_{z,A}=0}^{k_{\max}} v_{z,A} \hbar \omega \tau_{AB} \frac{df(\omega,T)}{dT} \right], \quad (1)$$

where the summation is performed over different polarizations denoted by $p_A$ and allowed wave vectors $k_{x,y,z}$ in material $A$; $V_A$ is the volume of material $A$, $v_{z,A}$ is the phonon group velocity normal to the interface, $\hbar\omega$ denotes the phonon energy wherein $\hbar$ is Planck's constant divided by $2\pi$ and $\omega$ is the phonon frequency, $\tau$ is the phonon transmission probability, $f$ is the phonon distribution function (Bose-Einstein distribution). The different PGM based formalisms differ based on how each method calculates the transmission probability. For calculating the thermal conductivity of different materials, PGM based formalisms have achieved excellent agreement with experimental measurements [ref]. When it comes to $G$, however, no general or consistent agreement between theory and experiment has ever been reported. Figure 1 compares the predictions of various theoretical formalisms (vertical axis) with the experimentally measured values (horizontal axis).

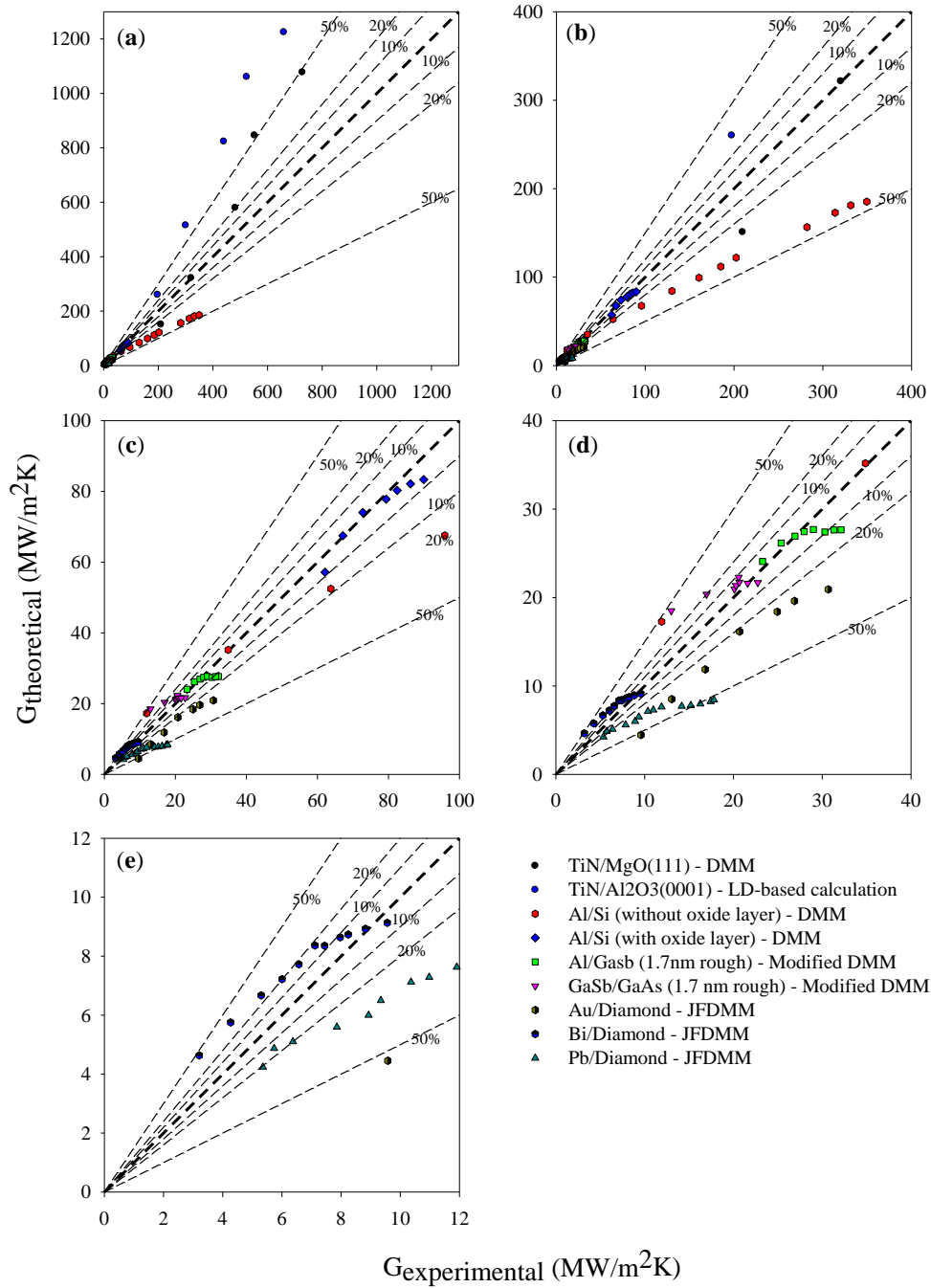

Fig.1. Comparison of theoretical predictions of thermal interfacial conductance, *G* across different interfaces with corresponding experimentally measured values. For each interface, each point represents a calculation/measurement at a different temperature. Each panel represents a different $G$ range. The dashed lines represent the percent error associated with the theoretical predictions. The examined interfaces have references as follows: TiN/MgO,[29] TiN/Al2O3,[29] Al/Si with/without oxide layer,[30] Al/GaSb,[31] GaSb/GaAs,[31] Au/Diamond,[32] Bi/Diamond,[9] and Pb/Diamond.[9] Modified DMM referenced in the legend was proposed as a variation of DMM to predict TIC across interfaces with severe chemical and structural changes around the interface.[33,34] JFDMM is a variation of DMM, where the altered phonon frequencies in the interface region is also included in the calculations.[35]

It is apparent from Fig.1 that all the PGM models suffer significant qualitative and quantitative uncertainty in their predictions. Note that in the Landauer formalism (eq. 1), group velocity, $v$ of all vibrational modes need to be calculated to evaluate $G$. Calculation of $v$ is only possible for purely crystalline solids. When PGM is applied to calculate interfacial conductance, it assumes the interface to be merely a break in symmetry of a crystal and adopts the same treatment used in perfect crystals. Such an assumption in applying PGM to systems with break in symmetry such as interfaces and disordered materials like amorphous materials, alloys, and polymers is highly questionable, since a large population of the vibrational modes in these systems are non-propagating and localized, for which a group velocity cannot be defined. In addition, PGM formalisms do not include the intrinsic anharmonicity of vibrations arising from the differences in modal frequencies, which is another important factor that affects heat conduction. Although Mingo has shown that, in principle, anharmonicity can be included, this has been neither widely adopted nor applied to any realistic interface. Finally, from eq. (1), it is clear that PGM defines $G$ by using the properties of only one of the two materials forming the interface and the transmission probability, $\tau$. AMM and DMM methods calculate transmission probability under the assumption of purely specular and purely diffuse scattering of phonons respectively at the interface, which are severely restrictive whatsoever. Although there has been many improvements made to these models from a standpoint of calculating transmission probability, none of them could resolve atomic level detail of interface quality like imperfections, defects, interatomic diffusion, etc.

Among the PGM based models, AGF formalism seems superior as it is capable of including atomic level details and quantum effects. AGF utilizes the harmonic force constants (FC) estimated by means of atomic forces calculated through first principle methods like density functional theory (DFT) or by empirical interatomic potentials to estimate the transmission function. In AGF method, for contact area $A$, $G$ is given by,

$$G = \frac{1}{2\pi A} \int \hbar\omega \frac{\partial f}{\partial T} \tau(\omega) d\omega, \qquad (2)$$

where the phonon transmission $\tau$ at frequency $\omega$ is calculated as the trace over the Green's function of the interface and its coupling terms between the bulk material on either end (explained in supplementary material). Although AGF, in theory, combines atomic-scale fidelity with an asymptotic treatment of the bulk material, to our knowledge, no good agreement with experimental measurements have been reported till date. Note also that AGF is intrinsically unable to achieve mode-level details of conduction. Although one may argue that mode-level details can be extracted from the interatomic force constants, since AGF calculates contributions only from the modes existing in the bulk material, it is uncertain whether they are the actual modes present in the interfacial system. The contributions from individual modes are particularly important in materials where all the eigen modes may not be propagating. The non-propagating modes could be localized or diffusive in nature. Knowing the contributions of specific eigen modes facilitates rational design of materials by engineering certain features to target certain group of modes to either inhibit or enhance their role. Therefore, it is presumable that, irrespective of the underlying theory, emphasis should be given on describing the contributions from the actual modes that exist in the system.

Unequivocally, there is a major gap in the understanding of interfacial heat transport. Considering the inadequacies in PGM, we seek an alternative view of interfacial heat conduction based on the fluctuation dissipation theorem wherein the modal contributions to transport are assessed by the degree to which they are correlated, rather than the degree to which they are scattered. To serve the purpose, the recently reported interfacial conductance modal analysis (ICMA) formalism[36] based on the fluctuation-dissipation theorem and lattice dynamics looks promising. In the ICMA formalism, the instantaneous energy transfer across an interface of material A and B can be given as:

$$Q_{A \to B} = -\sum_{i \in A} \sum_{j \in B} \left\{ \frac{p_{i,\alpha}}{m_i} \left( \frac{-\partial H_j}{\partial \mathbf{r}_i} \right) + \frac{p_{j,\alpha}}{m_j} \left( \frac{\partial H_i}{\partial \mathbf{r}_j} \right) \right\}. \quad (3)$$

Here, $Q_{A \to B}$ is the instantaneous energy transfer across the interface of material *A* and *B*; *p, H,* and *m* represent the momentum, Hamiltonian, and mass of atoms *i* and *j* respectively. From this relation, the conductance can be calculated by the time integration of correlation of autocorrelation of the equilibrium fluctuations of the heat flow as:

$$G = \frac{1}{Ak_B T^2} \int_0^\infty \langle Q_{A \to B}(t) \cdot Q_{A \to B}(0) \rangle dt \quad (4)$$

Since ICMA is implemented in classical molecular dynamics (MD) framework, it is capable of full inclusion of anharmonic contributions to the interfacial heat transfer by all types of vibrational modes including the localized interfacial modes. Most importantly, ICMA can resolve the modal heat flux $Q_n$ (i.e. $Q = \sum_n Q_n$) yielding the modal contribution to conductance, $G_n$ (i.e. $G = \sum_n G_n$) by utilizing the input eigen vector basis set given as input (detailed formalism given in supplementary material).

In this work, we report the first ever accurate theoretical prediction of thermal conductance for any interface using ICMA. We implement both ICMA and AGF techniques independently to predict thermal conductance of aluminum (Al)-sapphire (α-$Al_2O_3$) interface for temperature ranging from 50-500 K and compare the results with each other and with experiments. $Al/Al_2O_3$ interface is of high technological importance in space and underwater propulsion applications. Composite energetic materials synthesized with nanosized aluminum structures are promising high-energy density propellants. Quantifying $Al/Al_2O_3$ interfacial conductance is crucial in precise-modeling of combustion of these materials whose commercialization has so far been impeded by the apparent high thermal resistance. Results are also compared with DMM predictions to observe how results vary when purely diffuse scattering mechanism is assumed. Moreover, we also report the mode-level details obtained from ICMA, which gives the nanoscale picture of the modal interactions and thereby explaining the mechanisms governing interfacial heat transport.

## 2. Simulation details

### 2.1. Interfacial Conductance Modal Analysis (ICMA)

We used ICMA method in equilibrium molecular dynamics (EMD). A simulation cell size of ~19.2 nm in length having a cross sectional area of ~80 $nm^2$ with 1260 atoms containing an interface with crystal directions (111) Al || (0001) $Al_2O_3$ representing the primary orientation in FCC metal-metal oxide (Medlin, pilania) interfaces is used for the simulations. The system length was chosen based on an initial size-dependency calculation, which suggested that a system size larger than ~18 nm yields a size independent $G$. To model atomic interactions, we have used the Streitz-Mintmire (SM) potential, a variable charge interatomic potential designed specifically for Al/Al$_2$O$_3$ interface, which explicitly includes variable charge transfer between anions and cations in the material. SM potential has been used to adequately describe the elastic properties, surface energies, and surface properties of sapphire in prior works. Here we show that SM potential is also able to accurately describe the phonon properties of the bulk of both materials (supplementary material), and is therefore suitable for ICMA calculations.

Firstly, the system was relaxed in isobaric-isothermal (NPT) ensemble at zero pressure for 2 ns to relieve any internal stresses. After relaxation, the system was equilibrated in a canonical (NVT) ensemble at required temperature for another 2 ns. Following equilibration, the system was evolved in time under microcanonical ensemble (NVE) ensemble for 10 ns. Heat flux was recorded very 5 fs, which is found to be sufficiently low enough to resolve the heat current fluctuations in both the materials. In order to overcome the possible statistical uncertainty due to insufficient phase space averaging, 10 independent ensembles are considered for each temperature. All calculations were performed on Large-scale Atomic/Molecular Massively Parallel Simulator (LAMMPS) package using a time step size of 1 fs. To include the modal decomposition routine, the original SM potential in LAMMPS was modified to accept eigenvector basis set obtained from lattice dynamics (LD) calculations and to output modal contributions to heat current at required intervals. For performing lattice dynamics (LD) calculations, following the NVT equilibration, the system was gradually cooled to 0 K in microcanonical ensemble using Langevin thermostat. The system was then allowed to undergo relaxation at 0 K in NPT ensemble. The relaxed crystal was used as the input for LD calculations performed on General Utility Lattice Program (GULP) from which the eigen vectors of vibration for the structure were obtained. The auto- and cross-correlations between the total and modal heat fluxes from ICMA routine were calculated to obtain the total $G$ and modal contributions, respectively.

*2.2.Atomistic Green Function (AGF)*

To ensure a fair comparison between AGF and ICMA, force constants used in AGF calculation were obtained from empirical LD calculations using the same Streitz-Mintmire potential used in ICMA calculations. In AGF method, the system consists of two bulk regions of aluminum and sapphire respectively, and an interface region of these two materials. For FC calculation, we used structures composed 36 and 60 atoms respectively to represent bulk structures of aluminum and sapphire, and 96 atoms for interface structure. LD calculations were performed using ALAMODE code. Several initial geometries with atoms displaced from their equilibrium coordinates were input to ALAMODE. To obtain empirical FC, ALAMODE was coupled with LAMMPS, and the SM potential was invoked to obtain forces acting on atoms corresponding to the displaced geometries. Based on the harmonic force constants obtained from SM potential, we solved eq. X-Y in the supplementary material to obtain phonon transmission and $G$.

3. **Results**

*3.1.Total conductance, G*

The total $G$ values as a function of temperature obtained from various sources are summarized in Fig. 2. Results of our ICMA and AGF calculations are compared with the experimental results from three different sources[11,32,37] as well as the DMM predictions.[11] Experimental data from 50-300 K are obtained from Stoner and Maris,[32] whereas the experimental results for 300-480 K are obtained from Hopkins et al.[37] Another set of experimental data and DMM calculations reported by Hopkins et al.[11] are also overlaid for comparison. Experimental data above 480 K is not available.

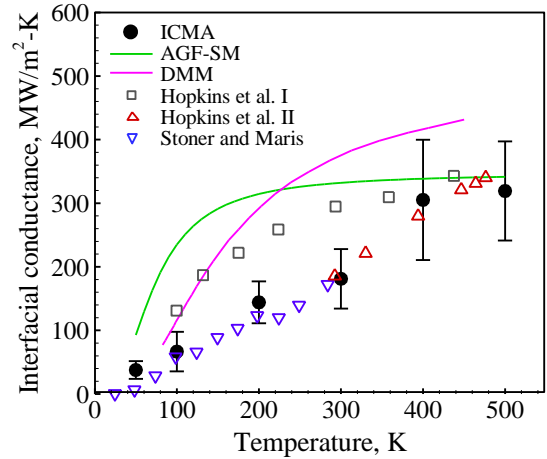

Fig. 2. Thermal interfacial conductance predicted by ICMA and AGF methods compared with experimental results and diffused-mismatch model (DMM) prediction

From Fig. 2, it can be seen that the ICMA results are in great agreement with experimental results for the entire range of temperature. The mean quantitative error in comparison with experiments is under 5% and the qualitative trend is also exceptional. Experimental results suggest a near-linearly increasing trend in $G$ *v/s* $T$ that is captured quite well by ICMA. As seen in Fig. 2, except for a slight proximity with Hopkins et al. experimental results below 150 K, DMM predictions are way above experimental results. This is because of the severely restrictive assumption on phonon scattering to be purely diffusive in nature. The green line represents the results of AGF calculations. It is quite evident that except for the bare accordance with the experimental result of Hopkins et al. at T = 450 K, AGF significantly over-predicts $G$ from 50-450 K. The temperature trend of $G$ predicted by AGF is also not in agreement with experiments. AGF results steeply rise from 50-100 K and plateaus thereafter. At higher temperatures, $G$ is nearly constant; no temperature dependence is observed. We doubt the poor predictability of AGF is a consequence of two factors. Firstly, AGF does not take into account the intrinsic anharmonicity associated with the vibrational modes, which plays a vital role in heat conduction. Secondly, the only mechanism that the AGF accounts for, in evaluating the temperature dependence is the quantum (Bose-Einstein) correction applied to the modal calculations. This is different from ICMA in which we calculate conductance using classical MD followed by quantum correction at every single temperature, thereby circumventing the limiting assumptions of AGF in assessing temperature dependence. To examine these conjectures, a detailed analysis of modal anharmonic energy distribution and modal contribution to $G$ is performed in the following sections.

*3.2. Anharmonicity*

In order to quantify anharmonicity and its ramifications on $G$ prediction, we evaluate the anharmonic energy belonging to each mode of vibration in the interfacial system. In essence, harmonic and full potential energy of each eigen mode is calculated by the procedure previously reported by Gordiz and Henry [ref]. Harmonic energy, $\Omega_{i,n}$ attributed to atom $i$ by the $n^{th}$ eigen mode can be given as:

$$\Omega_{i,n} = \frac{X_{i,n}}{\sum_{i'} X_{i',n}} k_B T, \qquad (5)$$

where $X_{i,n} = \sqrt{\frac{m_i}{N}} \mathbf{e}_{n,i} \cdot \mathbf{e}_{n,i'}$, and $N$ is the total number of unit cells in the system, $m_i$ is the mass of atom $i$, and $\mathbf{e}_{n,i}$ is the eigenvector associated with atom $i$ participating in eigen mode $n$. Utilizing the exact same displacements for a singly excited mode in the system, we can evaluate the total potential energy for each atom in the system for that eigenmode, $\Phi_{i,n}$. This is achieved by evaluating the total potential energy of the system as the summation over the individual atomic potential energies as:[38,39]

$$\Phi = \sum_i \Phi_i, \qquad (6)$$

where $\Phi$ is the total potential energy and $\Phi_i$ is the potential energy assigned to atom $i$, such that all energy is equally partitioned amongst interacting pairs of atoms. For an eigenmode $n$, the difference between the total potential energy $\Phi_{i,n}$ and the harmonic potential energy $\Omega_{i,n}$ associated with atom $i$ equals the anharmonic portion of the energy, $\Upsilon_{i,n}$ given by:

$$\Upsilon_{i,n} = \Phi_{i,n} - \Omega_{i,n} \qquad (7)$$

By summing $\Upsilon_{i,n}$ over atoms, we can inspect the anharmonic energy belonging to the vibrational modes in the system. This can then be used to better understand how modes and regions of atoms interact and ultimately will help to quantify its effect on transport. Figure 3(a) shows the mode level degree of anharmonicity by means of two representative plots: a) anharmonic energy normalized by $k_BT$ and b) $G$ accumulation function normalized by total $G$, both at T = 300 K. From Fig. 3 (a), it can be seen that the normalized anharmonic energy clusters around zero for frequencies less than ~9 THz, which is approximately the maximum frequency of vibrations in the aluminum crystal, and severe anharmonic behavior is observed for frequencies > 9 THz. This knowledge may be used to roughly assess the fidelity of any theoretical model which assumes purely harmonic vibrations. In other words, if majority of modes fall below 9 THz, harmonic assumption to vibrational modes may be reasonable and AGF may, therefore, be expected to predict $G$ with reasonable levels of accuracy.

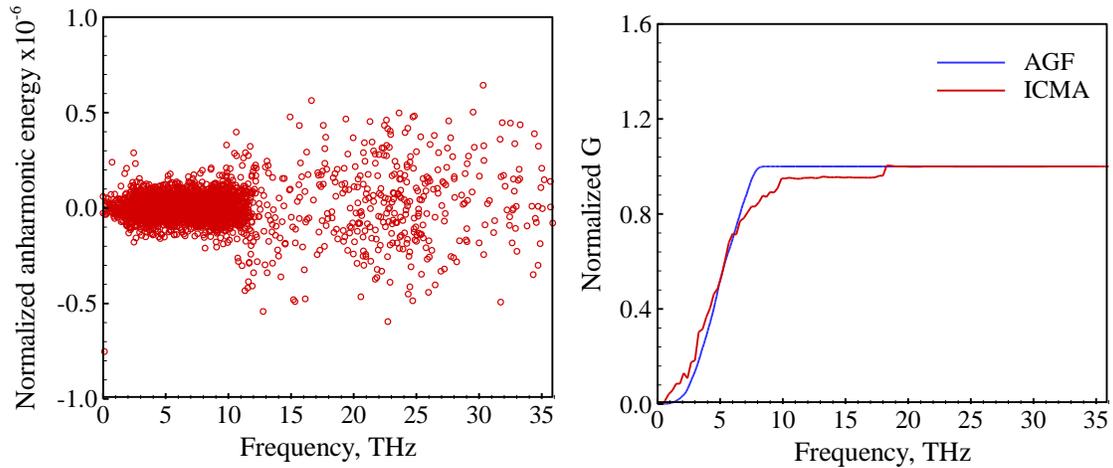

Figure 3 (a) Mode level distribution of anharmonic energy normalized by $k_BT$ clearly showing the increased anharmonic behavior of modes with frequency > 9 THz and b) Normalized conductance accumulation plots obtained from AGF and ICMA calculations clearly showing the inability of AGF to account for the intrinsic anharmonicity within the interfacial system, both at T=300 K.

On further observing the normalized conductance accumulation function ($G(\omega)/G$) as shown in Fig. 3(b), the effect of anharmonicity is clearly visible. AGF results accumulate by around 8 THz and stays constant thereafter. It is, therefore, unable to capture contributions from modes with frequencies greater than the highest modal frequency in aluminum. On the other hand, ICMA trails a steep rise until 8 THz followed by a reduced steepness until 10 THz then staying constant until 17.5 THz. The reduced steepness is because of the onset of anharmonic effects for modal frequencies > 9 THz. At this point, it is worthwhile looking at the distribution of different types of modes in the system to assess whether ICMA results align well with experimental results for the right physical reasons.

*3.3.Modal Analysis*

Phonon DOS of bulk aluminum and aluminum oxide, partial DOS, modal summation of TIC, and mode-mode correlations at 300 K is shown in Fig. 4 (a)-(d) respectively. Fig. 4 (a) shows the DOS of bulk materials computed from the Fourier transform of velocity autocorrelation function obtained from MD calculations. Partial DOS shown in Fig. 4 (b) are calculated from the eigenvector basis set obtained by lattice dynamics calculations. From the partial DOS, we can identify four types of modes based on their participation ratio as: i) extended modes, ii) partially extended modes, iii) isolated modes, and iii) interfacial modes. Extended modes are present at the interface, but majority of them is not at the interface, and are delocalized into both materials. Partially extended modes are also present at the interface, but majority of them are not present at the interface, and are localized on one side of the interface. Isolated modes exist far away from the interface while interfacial modes are localized vibrational modes which are majorly present at the interface. Figure 4(b) shows the dominance of partially extended modes in Al (< 9 THz) and $Al_2O_3$ (<12 THz) and the negligible presence of extended modes (< 0.6 THz). A small percentage of modes (> 9 THz) are interfacial in nature whereas the remaining modes seem to be isolated.

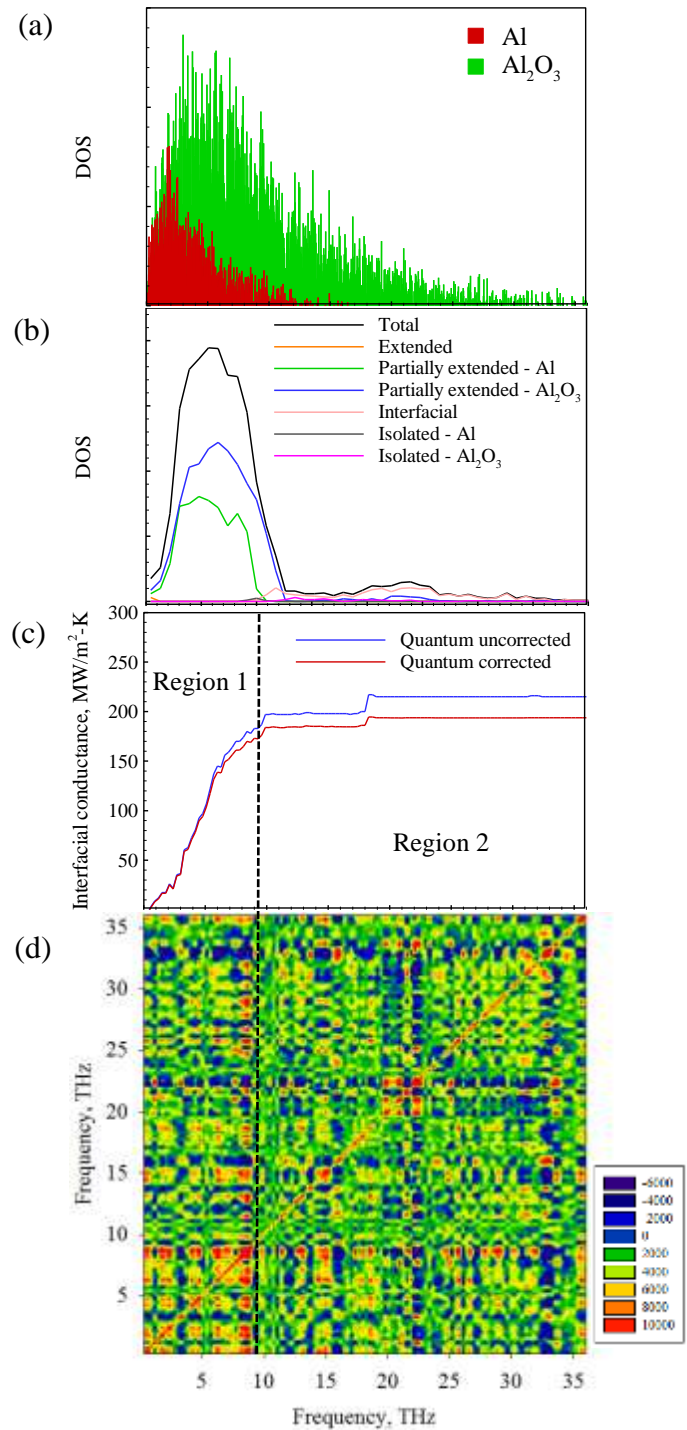

Fig. 3. a) Bulk DOS of Al and $Al_2O_3$ obtained from the velocity autocorrelation function b) Partial DOS showing different types of modes, c) TIC accumulation (In this system, >90% of the total conductance is contributed by partially extended modes on Al and Al2O3, and the extended modes below 12 THz, d) mode-mode correlation map at T = 300 K showing three distinguishable regions. Region 1 is mostly comprised of partially extended modes on Al and $Al_2O_3$; this region ends with the maximum frequency of partially extended modes and contributes >90% to TIC. Region 2 has almost negligible contribution to TIC and are mostly comprised of interfacial modes, which explains the reason behind their low contribution. The frequency range 17.5-22.5 THz, however, has a high density of partially extended and interfacial modes in $Al_2O_3$ co-existing that interact and contribute ~5% to TIC.

In Fig. 4(c) and (d), two distinguishable regions are marked as Region 1 and 2. Region 1 is below 9 THz marking the peak frequency in Al. Region 2 comprises all frequencies above 9 THz. In region 1, distinct regions of strong positive correlations are observed. Except below 0.6 THz where extended modes are present, this region is dominated by the cross-correlation (CC) of partially extended modes of Al with that of $Al_2O_3$ reflecting as red regions in the correlation map. Considering the large population of states of partially extended modes of Al and that of $Al_2O_3$, this region of high CC is the major reason for the initial high slope of TIC accumulation until ~ 9 THz. After that, within region 1, 9-12 THz marks a narrow region of interaction between interfacial and partially extended modes of $Al_2O_3$, which due to a combined effect of weak correlations and low density, only gives a very shallow slope in TIC accumulation.

From 12-15.5 THz, interfacial modes, and the isolated and partially extended modes of $Al_2O_3$ co-exist. This frequency ranges in region 2 also shows a strong correlation. However, the very low density of these modes is a clear evidence of the small increment of *G* in this region. In region 2, the modal characteristics shifting drastically from strong positively to strong negatively correlated regions in effect canceling each other, maintain cumulative $G$ constant until ~17.5 THz. From 17.5 THz to 22.5 THz, there are observable regions of strong correlation and a slight increase in the density of vibrational states. Especially around the diagonal, there is a strong observable positive correlation from the interaction between the interfacial modes. The combined effect is a jump in $G$ accumulation in Fig. 4 (c) at around the same frequency i.e. 17.5 THz. In order to further gauge the role of each mode, it is important to obtain the contribution of each type of mode to *DOS* and $G$ and the relative contribution of each mode to $G$ i.e. *G/DOS*.

Table 1. Contribution of different types of modes to partial DOS and G, and the percentage relative contribution of G to DOS

| Mode Type | $DOS(\%)$ | $G(\%)$ | $G/DOS$ |
|---|---|---|---|
| Extended | 0.18 | 0.31 | 1.72 |
| Partially extended | 89.52 | 91.83 | 1.02 |
| Interfacial | 9.31 | 7.80 | 0.84 |
| Isolated | 0.90 | 0.06 | 0.07 |

Table 1 shows the population of each type of mode in partial *DOS*, and their contribution towards $G$. Also given is the percentage relative contribution of $G$ to *DOS* to comprehend the significance of each type in conducting heat. The high density of partially extended modes in the region from frequency < 12 THz corresponding to 89.52% of DOS together constitute towards ~92% of TIC. The *G/DOS* ratio of partially extended modes ~1 suggesting that the role of partially extended modes in *G* is justified by their presence in the partial *DOS*. In the remaining 10.5% modes, 9.31% is constituted by interfacial modes and under 1% by isolated modes. The percentage of extended modes in the system is only 0.18%. Considering the *G/DOS* value of 1.72, it is to be understood that there is a disproportionately high contribution to *G* from the extended modes for their relatively small presence in *DOS*. Therefore, we doubt that the presence of high concentration of extended modes is the reason for the apparent high *G* of cSi-cGe[40] interfacial system reported in a prior work. We believe that for the Al/Al$_2$O$_3$ system, a major portion of *G* is contributed by partially extended modes, subsequently helping us achieve realistic predictions. In summary, ICMA has not only been able to provide an accurate theoretical prediction of interfacial conductance, but also capture the physical picture of modal interactions governing thermal transport.

## 4. Conclusions

We have accurately predicted the thermal interfacial conductance (*G*) of the aluminum (Al)-sapphire (α-Al$_2$O$_3$) interface along the crystal directions (111) Al ∥ (0001) Al$_2$O$_3$. We used two fundamentally different formalisms: interfacial conductance modal analysis (ICMA) and atomistic green function (AGF) method in the temperature range 50-500 K. While AGF overpredicts *G*, predictions of ICMA show great agreement with experimental results both quantitatively and qualitatively. This is for the first time ever in literature, a theoretical model has been able to predict thermal conductance of a realistic interface for a wide range of temperatures and achieve a conclusive experimental validation. ICMA formalism is clearly superior to all the

PGM based models for its more fundamental treatment of the interfacial heat flux, its inclusion of full anharmoncity of vibrational modes, and for its ability to access to phonon mode level details. High thermal resistance of Al/Al$_2$O$_3$ interface has been a major impeding factor in the massive commercialization of nano-aluminum based energetic materials. These results are of vital importance to the heat transfer and space propulsion community; a major milestone in combustion heat transfer, which would enable mars propulsion and *planet hopping* a reality in the near future.